\documentclass{article}
\usepackage[left=3cm, right=3cm, top=3cm, bottom=3cm]{geometry}
\usepackage{graphicx}
\usepackage{authblk}
\usepackage{amsmath}
\usepackage{mathrsfs}
\usepackage{amsfonts}
\usepackage{dsfont}
\usepackage{gensymb}
\usepackage{hyperref}
\usepackage{MathUnicode}
\usepackage{xcolor}
\usepackage{pagecolor,lipsum}% http://ctan.org/pkg/{pagecolor,lipsum}

\definecolor{bookColor}{cmyk}{0	, 0  , 0   , 0.00}  % 0.90\% of black
\color{bookColor}

\DeclareFontFamily{U}{wncy}{}
\DeclareFontShape{U}{wncy}{m}{n}{<->wncyr10}{}
\DeclareSymbolFont{mcy}{U}{wncy}{m}{n}
\DeclareMathSymbol{\Sh}{\mathord}{mcy}{"58}

\newcommand{\rmd}{\textrm{d}}

\title{Resolving Exo-Continents with Einstein Ring Deconvolution}
\author{Alexander Madurowicz}
\date{April 1, 2020}
\affil{Kavli Institute for Particle Astrophysics and Cosmology, Stanford University}
\begin{document}
\pagecolor{black}
\maketitle
\begin{abstract}
    A mission to the focus of the solar gravitational lens could produce images with unprecedented angular resolution and sensitivity. In the context of trying to resolve the time variable thermal signature of continents on other Earth-like exoplanets, we develop an approach to improve the image reconstruction performance by using azimuthal variations in the Einstein Ring's intensity. In the first post-Newtonian approximation to General Relativity, an arbitrary disk intensity distribution in the source plane is mapped to a narrow annulus around the Einstein Ring, with each azimuthal element corresponding to a sector in the disk. A matrix-based linear measurement model at various fixed signal-to-noise ratios demonstrates that this extra information is useful in improving the reconstruction when the image is sparsely sampled, which could improve integration times and temporal errors. Various issues and future outlooks are discussed.
\end{abstract}
\section*{Introduction}
Let us suppose for a moment that the year is 2050. The past few decades have been a boon for the field of exoplanetology. With the successful completion of the LUVOIR mission \cite{luvoir}, many Earth-like exoplanets have been discovered and characterized in nearby systems, with a select few demonstrating potentially interesting biomarkers. You and your science team have published a great number of truly remarkable papers on these discoveries, but are sick and tired of including artist's renditions of exoplanets in your press releases. A spectrogram simply does not capture the public interest. You and others around the world are clamoring for a real image of another one of these worlds, with all of the beautiful detail akin to the iconic Blue Marble image of our home world taken by astronauts aboard the Apollo spacecraft. What would it take to obtain such an image?

You pull out a napkin to do an order of magnitude calculation of this absurdity. You know the Earth is about 6000 km in radius, and a typical system is about 100 pc away. So if you want to resolve 1000 pixels across its surface, you estimate the typical angular scale of such an image would be $10^{-15}$ radians. Since the resolution of your telescope is about $\lambda/D$, at optical wavelengths around 500 nm, you find you would need to build a telescope with diameter $D = 250,000$ km! That's nearly as big as the Sun! Congress would never approve that in the budget.

In your pondering of the absolute insanity that would be a telescope the size of the Sun, you stumble upon a rather interesting idea. Why not use the Sun as a gravitational lens? You do a bit of digging, and find that it \cite{Landis2016} has \cite{Turyshev2019} already \cite{alkalai2017} been \cite{Turyshev2018} thought \cite{Turyshev2018b} of \cite{MACCONE2011}, even since before you were born \cite{ESHLEMAN}. Some people just get an unfair headstart. 

\begin{figure}[h!]
    \centering
    \includegraphics[width=\textwidth]{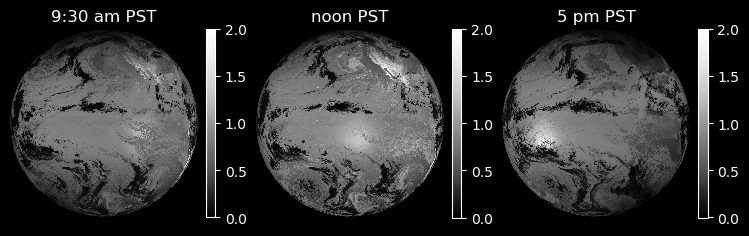}
    \caption{Level-1b radiances from the GOES-R weather satellite \cite{goesr} in the shortwave window at 3.9 $\mu$m. The thermal response of the North America continent to the changing solar insolation is visible on a daily timescale. In addition, the specular reflection off of the surface of the Pacific ocean, and other changes due to atmospheric variability are visible.}
    \label{fig:time}
\end{figure}

In this paper we propose a different measurement paradigm from what has been previously \cite{Turyshev2020} explored. Instead of using the entire telescope as a single pixel detector, measuring the Einstein Ring's total brightness at various locations, we propose an instrument concept which could measure the azimuthal variations in the intensity of the Einstein Ring. This additional information could improve the reconstruction performance for a sparsely sampled and time-variable image. For an example science case, consider the image in Figure \ref{fig:time}, which shows the Earth from the view of a geostationary weather satellite. Over the course of a day, the thermal response of the North American continent is evident, as the land mass rapidly heats and cools in response to the variable insolation. The image changes subtly in many other ways, which all complicate a reconstruction that would require a long integration time.

The structure of the paper is as follows. We begin by introducing the necessary background physics behind gravitational lensing, which leads into a discussion of the azimuthal structure of the Einstein Ring. In the second half, we build a matrix based approach to simulating these measurements and demonstrate that a reconstructor which uses the azimuthal information is superior to one which does not, for an equivalent signal-to-noise ratio on each type of measurement. The paper conclude by broadly discussing issues surrounding this mission concept, and potential outlooks for its future.

\section*{Gravitational Lensing}
In the first post-Newtonian approximation to General Relativity, the metric for a spacetime with a Newtonian gravitational potential $Φ$ satisfying Poisson's equation $\nabla^2Φ = 4πGρ$  is \cite{carroll2003}
\begin{equation}
    \rmd s^2 = -\Big(1 + \frac{2Φ}{c^2}\Big)\rmd t^2 + \Big(1 - \frac{2Φ}{c^2}\Big)\Big(\rmd x^2 + \rmd y^2  + \rmd z^2\Big),
\end{equation}
and the photon trajectories associated with the corresponding null geodesics can be found by applying Fermat's principle of least time to a spacetime with a globally variable index of refraction given by \cite{Schneider1992}
\begin{equation}
    n = 1 + \frac{2}{c^2}|Φ| .
\end{equation}
As a consequence, the deflection angle $\vec{α}$ is given by the integral along the path of the gradient of the index of refraction in the direction perpendicular to the photon path  \cite{Narayan1996}
\begin{equation}
    \vec{α} = - \int \vec{\nabla}_{\perp} n \rmd l.
\end{equation}
Assuming the source and observer rest at opposite infinities, and that the deflection is small, so we can compute the gradient along the unperturbed path, a point mass $M$ with a corresponding potential $Φ = -\frac{GM}{r}$ has a deflection angle which depends on the impact parameter $b$ of the incoming photon:
\begin{equation}
    \vec{α} = \frac{2R_S}{b}\hat{b},
\end{equation}
where $R_S = \frac{2GM}{c^2}$ is the Schwarzschild radius. 

In principle it is possible to solve for the deflection angle for an arbitrary mass distribution by integrating the point source deflection for an infinitesimal mass element over an arbitrary density and volume, but for simplicity we make two approximations, the thin screen projection and axial symmetry, so that the projected density profile only depends on the radial impact parameter. These assumptions are fair because the solar oblateness is small, measured by precise instruments \cite{damiani2011} to be of order $\sim10$ milliarcseconds for the photosphere. Since the solar density profile rises steeply in the interior \cite{Dalsgaard1286} where the bulk of the mass is, deviations from axial symmetry are even less. The end result is that the Sun's three dimensional mass distribution lenses very nearly as a point mass would, which simplifies the further analysis.\footnote{We could have arrived at this same conclusion by assuming the Sun is perfectly spherically symmetric and noting that Birkhoff's theorem \cite{birkhoff} shows that the Schwarzschild solution is the only spherically symmetric vacuum solution to Einstein's equations, regardless if the mass distribution is a point or a uniform density sphere or any other spherically symmetric density distribution.} However, considering the violations of this assumption are likely important, as demonstrated by Loutsenko \cite{Loutsenko2018} and should be considered in a more careful treatment.

\begin{figure}[h!]
    \centering
    \includegraphics[width=.7\textwidth]{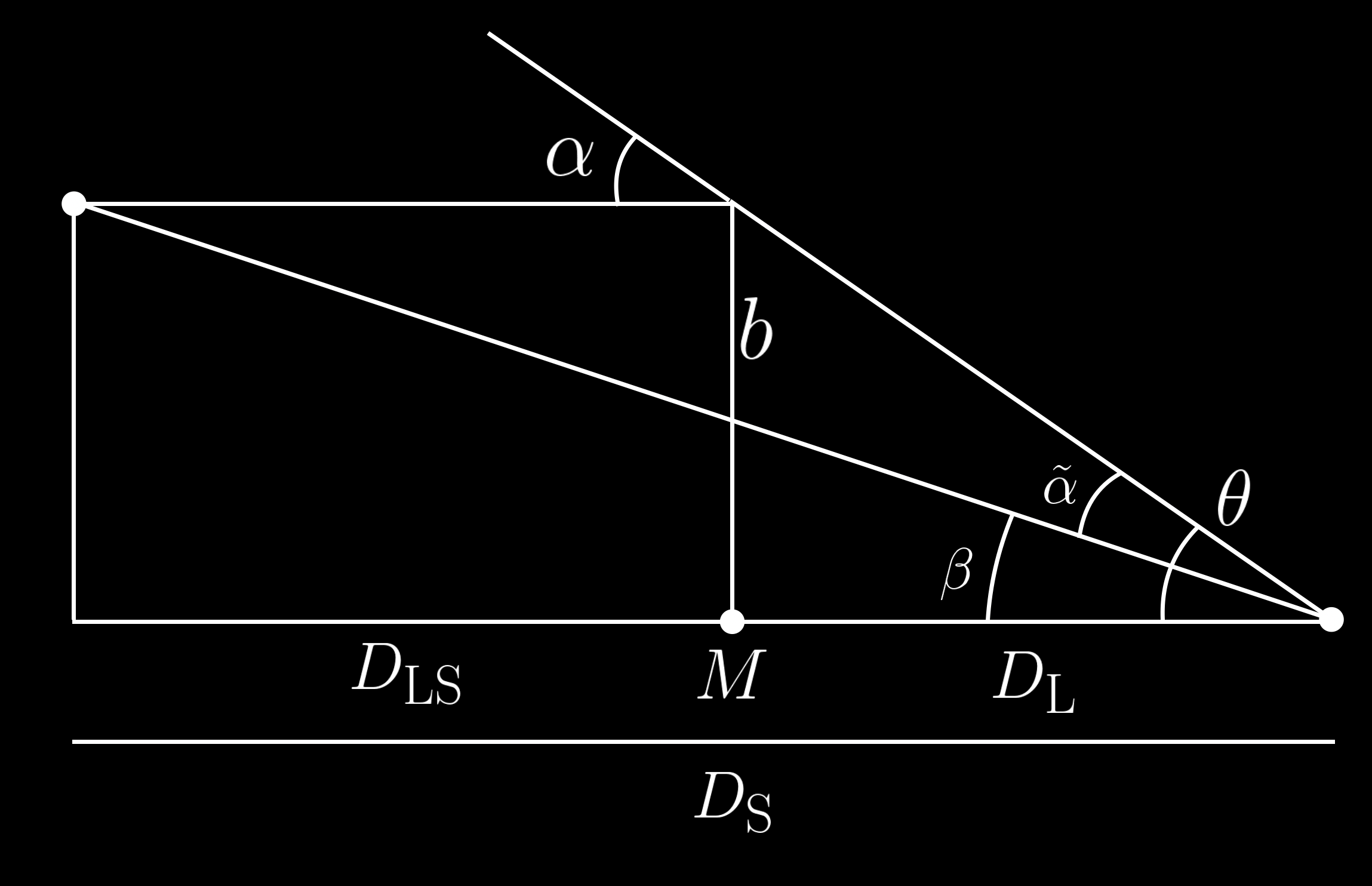}
    \caption{Lensing Geometry.}
    \label{fig:lens}
\end{figure}

Under our simplifying assumptions, it is rather straightforward to solve for the effect of a gravitational lens on a point source of light, and therefore an arbitrary extended source by means of superposition. By introducing a generic geometry for the lensing problem in Figure \ref{fig:lens}, and defining the reduced deflection angle
\begin{equation}
    \tilde{α} = \frac{D_{\textrm{LS}}}{D_\textrm{L}}α,
\end{equation}
the relationship between a point source's true and apparent locations $β$ and $θ$ is rather straightforward
\begin{equation}
    \vec{θ} = \vec{β} + \vec{\tilde{α}}(\vec{θ}).
\end{equation}
This is known as the lens equation.

\begin{figure}[h!]
    \centering
    \includegraphics[width=.6\textwidth]{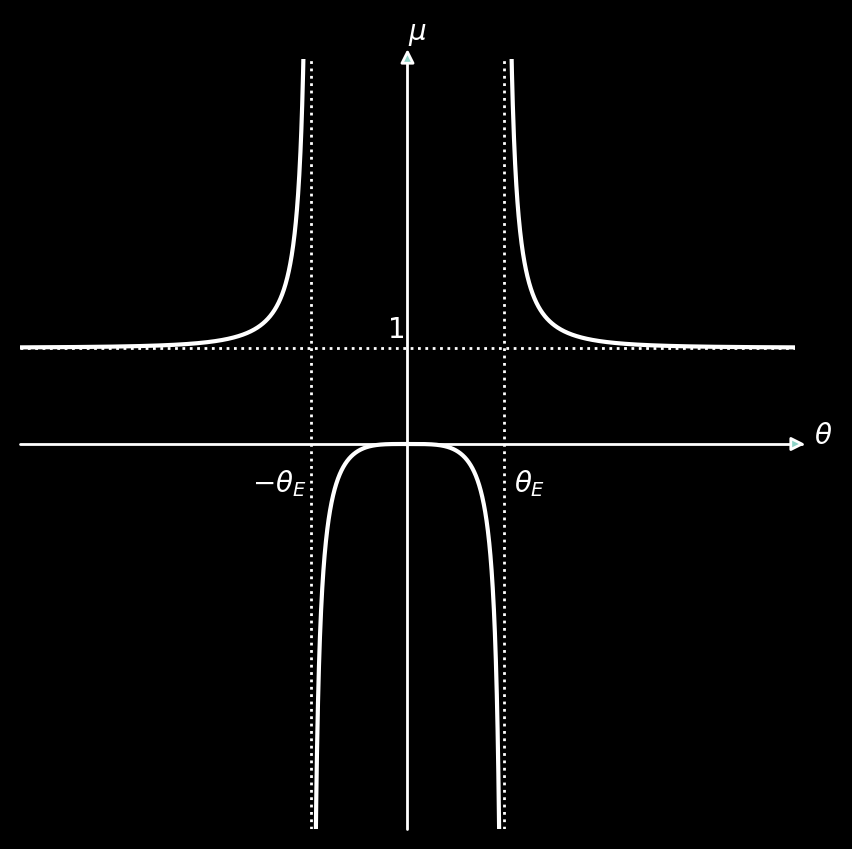}
    \caption{Magnification of a point source lens}
    \label{fig:mu}
\end{figure}

The lens equation can be solved for the point mass deflector, and the solution generically produces two results 
\begin{equation}
    θ_\pm = \frac{1}{2}\Big(β \pm \sqrt{β^2 + 4θ_E^2}\Big),
\end{equation}
where $θ_E$ is known as the Einstein Ring radius and is the solution when $β=0$,
\begin{equation}
    θ_E = \sqrt{2R_S\frac{D_{\textrm{LS}}}{D_{\textrm{S}}D_{\textrm{L}}}}.
\end{equation}
Additionally, the lens not only acts to modify the image location, but the image size as well. By defining the magnification as the ratio of the image area to the source area
\begin{equation}
    μ = \frac{θ}{β}\frac{\rmd θ}{\rmd β},
\end{equation}
we can find that a point source magnifies the images by
\begin{equation}
    μ = \Big[1 - \Big(\frac{θ_E}{θ_\pm}\Big)^4\Big]^{-1},
\end{equation}
which is discontinuous at infinity when $β=0$, see the plot in Figure \ref{fig:mu}. This discontinuity is the result of the lens mapping a measure zero subset of the source plane into a different measure zero subset of the lens plane. More specifically, the single point of the source which is colinear with both the observer and the point lens is mapped into a ring of points in the lens plane at an angular location corresponding to the Einstein radius. Since both of these subsets are measure zero, with no area, the magnification is ill-defined, essentially 0/0. This does not mean that the Einstein ring is infinitely bright. Misner, Thorne and Wheeler have shown that a version of Louisville's Theorem of conservation of phase space volumes, when applied to gravitational lenses, preserves the specific intensity of light rays along null geodesics (up to a factor of frequency shift due to the relative gravitational redshift along the path). \cite{Misner1973} Because the specific intensity is conserved along the light rays, and only a finite number of photons are generated at the source, only a finite number of photons arrive at the observer. 

\section*{The Azimuthal Profile of the Einstein Ring}

Let us consider the imaging of an extrasolar planet through the solar gravitational lens more concretely. Assuming the planet in question is roughly Earth-sized, with a radius $R_E \sim 6000$ km, in a system at a distance $D_{\textrm{LS}} = 100$ pc, and our observer is a roughly Hubble-sized telescope with diameter $D_{\textrm{tel}} = 1$ m at the heliocentric distance of $D_L = 600$ AU. Under these conditions, the angular scale of the Einstein Ring is $\theta_E = 1.67$ arcseconds. Since the radius of the Sun is $R_\textrm{Sun} \sim 7\times 10^8$ meters, it occupies an angular extent of $\theta_\textrm{Sun} = R_\textrm{Sun}/D_\textrm{L} = 1.59$ arcseconds, and the Einstein ring is just visible around the outer edge of the Sun. In the case where the planet, Sun, and telescope are aligned, the maximum value of $\beta$, which corresponds to the edge of the planet's disk, is $\beta_{\textrm{max}} = R_E/D_{\textrm{S}} \sim 2\times 10^{-12} $ radians or about $400$ nanoarcseconds. This results in an Einstein Ring width of $\Delta \theta = |\theta_\pm - \theta_E| \sim 200$ nanoarcseconds, which is essentially unresolveable for any conventional imaging system.\footnote{The Event Horizon Telescope, which recently produced the highest angular resolution image ever by means of a vast interferometric array distributed across the surface of the Earth \cite{EHT}, was able to image M87 at an angular resolution of roughly $20$ microarcseconds, which is two orders of magnitude larger.} So we can say that the Einstein Ring is essentially a ring, even though it is truly an annulus with a very narrow width. In principle an instrument with infinite resolution could resolve all of the features of the source using the radial variations of its intensity, in practice this is essentially impossible. However, useful and accessible information is still present in the azimuthal variations of the Einstein Ring's intensity, which can be resolved by our Hubble-sized instrument.
\begin{figure}[h!]
    \centering
    \includegraphics[width=.6\textwidth]{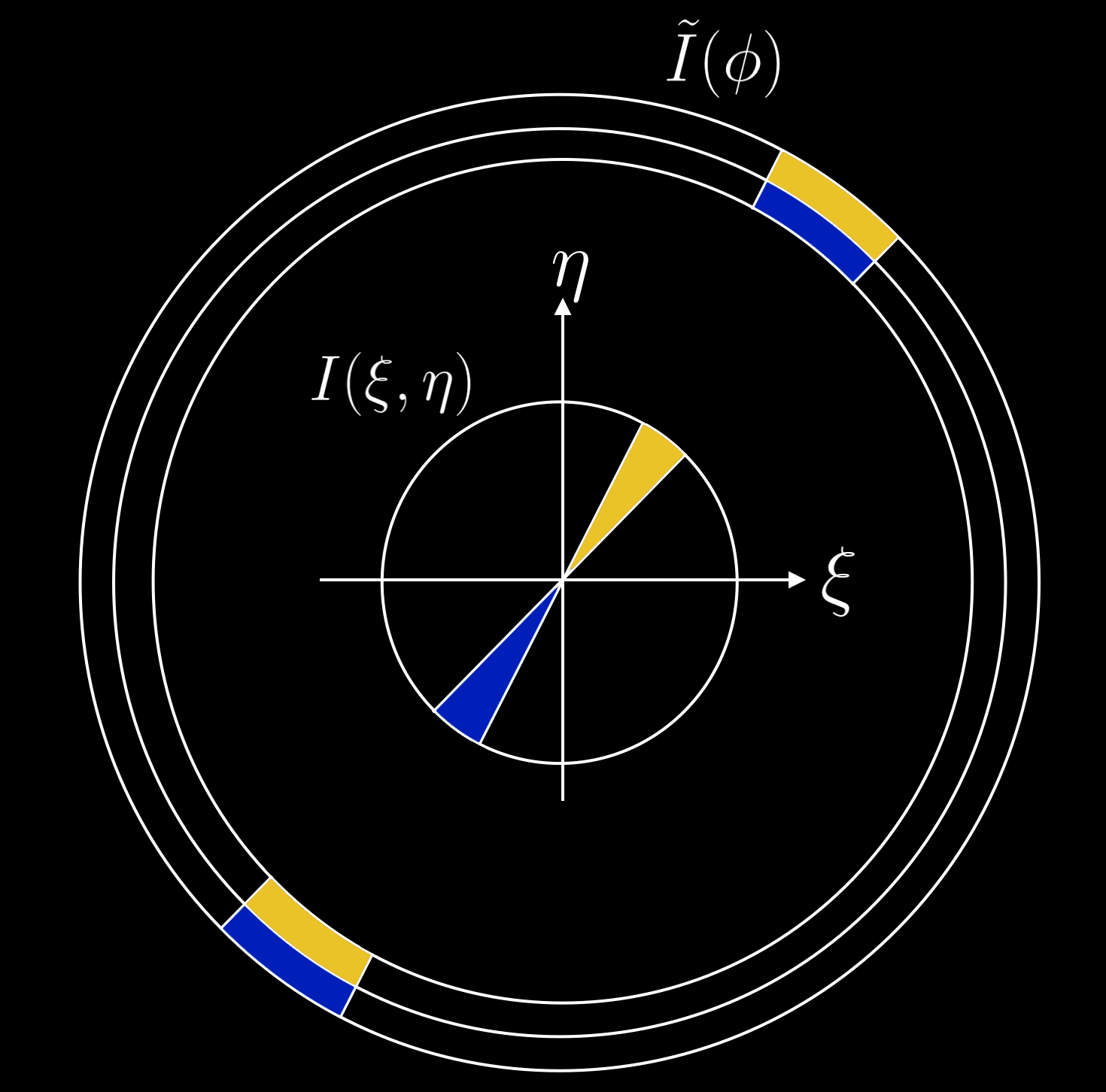}
    \caption{The correspondence between the location of intensities in the source plane and along the Einstein Ring for a singular azimuthal resolution element. The three concentric rings represent the center and edges of the Einstein Ring's annulus, at $\theta_E$ and $\theta_E \pm \Delta\theta$, respectively. }
    \label{fig:azimuthal}
\end{figure}

The forward model $M$ of the lensing problem for a generic two dimensional intensity distribution in the source plane $I(\xi,\eta)$ can be represented as a dimensional reduction to the one-dimensional Einstein Ring intensity profile $\tilde{I}(\phi)$
\begin{equation}
    I(\xi,\eta) \xrightarrow{M} \tilde{I}(\phi) .
\end{equation}
Since each point in the source plane creates two points $\theta_\pm$ in the image plane, one positively and one negatively magnified, on opposite sides of the point deflector, the azimuthal profile of the Einstein Ring has a spin-2 symmetry, where it would appear identical if rotated 180 degrees, like the face of a Jack, Queen, or King in a deck of cards. Each azimuthal element $\tilde{I}(\phi) \rmd \phi$ of the intensity profile is the sum of the intensities of the source plane whose azimuthal coordinate modulo $\pi$ corresponds to that particular value of $\phi$, multiplied by the absolute value of the magnification factor $|\mu|$. This is because negative magnifications simply invert the orientation of the image. This also assumes that the magnification factor is identical for opposing sides of the intensity distribution, but this is not strictly true. However, since $\log |\frac{\mu(\theta_+)}{\mu(\theta_-)}|\lesssim2\times10^{-7}$ for all of the relevant coordinates, this is a reasonable assumption. The final relationship between the input and output intensities can be expressed as
\begin{equation}
    \tilde{I}(\phi)\rmd \phi = \iint |\mu(\xi,\eta)| I(\xi,\eta) \mathcal{A}(\xi,\eta) \rmd \xi \rmd \eta,
\end{equation}
where $\mathcal{A}$ represents the azimuthal selection function
\begin{equation}
    \mathcal{A}(\xi,\eta) = | \arctan(\eta,\xi) \% \pi - \phi | < \frac{1}{2}\rmd\phi,
\end{equation}
which acts as a boolean aperture, taking on values of $0$ or $1$.

Using the Rayleigh resolution criterion, our instrument has angular resolution at a wavelength $\lambda = 1$ micron of $\theta_\textrm{tel} = 1.22 \lambda/D_\textrm{tel} \sim .25$ arcseconds, and since the Einstein Ring has a angular circumference of $\theta_\textrm{circum} = 2\pi\theta_E$, we can say that essentially our telescope can resolve $\theta_\textrm{circum} / \theta_\textrm{tel} \sim 42$ azimuthal resolution elements independently. If the telescope point spread function is well characterized for deconvolution or super resolution, it may be possible that more azimuthal elements can be measured, but this a useful starting approximation. 

\section*{Matrix Modeling}
The integral in equation (12) represents a linear transformation between the input and output intensity distributions $I(\xi,\eta)$ and $\tilde{I}(\phi).$ While it may be possible to simplify the aziumthal selection function by changing into polar coordinates, there is no reason we should expect the generic source intensity distribution to be analytic in any simple form. Since it arises from a combination of planetary albedo and solar astrophysics it would be likely be highly complex function that solves both the Navier-Stokes equations of hydrodynamics as well as the radiative transfer equations in three dimensions. This analysis is an interesting problem in its own regard, but for our purposes we wish to simplify the problem. By sampling the continuous intensity distribution $I(\xi,\eta)$ on a discrete grid $\Sh(\xi)\Sh(\eta)$, we can approximate the integral with a discrete sum
\begin{equation}
    \tilde{I}(\phi) \approx \sum_\xi\sum_\eta |\mu(\xi,\eta)| I(\xi,\eta) \mathcal{A}(\xi,\eta)\Sh(\xi)\Sh(\eta).
\end{equation}
Since the sum runs over both dimensions, if we wish to represent this sum with a single matrix operation, it will be necessary to unravel the indices, so that the discretely sampled $\mu$, $I$, and $\mathcal{A}$ all are represented as one dimensional vectors. If our two dimensional indices are $i, j \in 0,1,...,N-1$, where $N$ is an even integer box size for our simulation, then our discrete coordinate take on values
\begin{eqnarray}
    \xi_i = \frac{2i - (N-1)}{N}2\beta_\textrm{max}, \\
    \eta_j = \frac{2j - (N-1)}{N}2\beta_\textrm{max}.
\end{eqnarray}
This has the effect of displacing the centers of the pixels from zero by a half integer step, to avoid issues associated with dividing by zero and infinite magnification, and resizing the box dimension to correspond to coordinates in the range ($-2\beta_\textrm{max},2\beta_\textrm{max}$), so that the planet occupies half the total box dimension. When working with a high resolution dataset for the input intensity, it will be necessary to rescale and interpolate to match the grid size, but this is a standard feature in many imaging libraries.

\begin{figure}[h!]
    \centering
    \includegraphics[width=\textwidth]{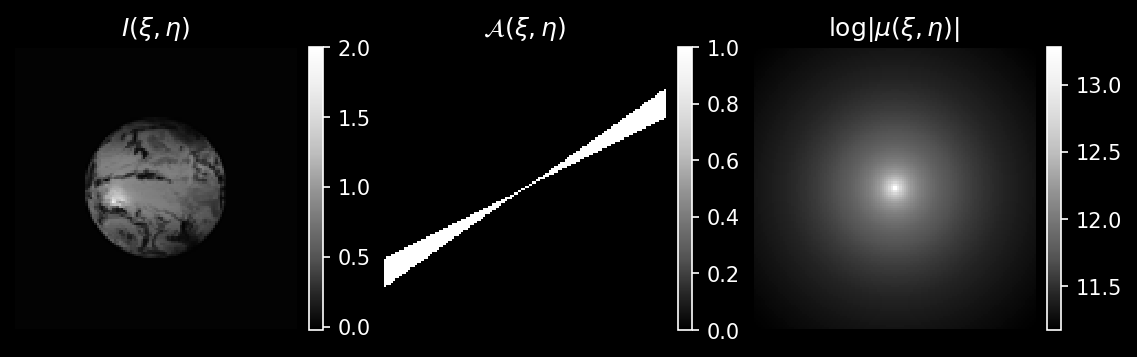}
    \caption{A simulated example of computing a singular aziumthal element using the discrete matrix approximation.}
    \label{fig:trip}
\end{figure}

The magnification can be computed directly from (7) and (10), using the notion that $\beta = \sqrt{\xi^2+\eta^2}$, and the azimuthal selection function can be computed from (13) for arbitrary values of $\phi$ and $\rmd \phi$. However, we find that in the discrete grid approximation, the azimuthal sampling becomes very poor near the origin, where only four pixels correspond to a rotation of $2\pi$, so we introduce a notion to correct for this. If the azimuthal distance given in (13) is less than $\frac{1}{2} \rmd \phi$ or it is less than
\begin{equation}
    \epsilon = \frac{1}{2\beta},
\end{equation}
then it is included in the boolean aperture. This can be thought of as encapsulating half the number of azimuthal radians per pixel at a certain angular separation $\beta$. At wide separations it is small and does not affect the aperture, but close to the origin it becomes large. This introduces lenience where the azimuthal sampling is small, resulting in more pixels being included near the origin, where the magnification is large and important.\footnote{This sort of computational issue highlights a type of interpolation problem when dealing with a polar symmetry on a square grid. A better approach could use a data structure better suited to polar coordinates, perhaps similar to how HEALpix \cite{healpix} operates on a sphere.} This process is demonstrated visually in Figure \ref{fig:trip}.

After all of the matrices corresponding to the two dimensional functions for $\mu$, $I$, and $\mathcal{A}$ are constructed, they can be unraveled into a vector following the convention of the last axis index incrementing most rapidly:
\begin{equation}
    k = j*N + i.
\end{equation}
This can be thought of as reading off the image pixels from left to right, top to bottom, and provides a mapping between our new one-dimensional index $k$ and the two dimensional indices $(i,j)$. The reverse mapping can be obtained with
\begin{eqnarray}
    i = k \% N \\
    j = k // N,
\end{eqnarray}
where $\%$ is the modulo operation and $//$ is the floor division operation, which throws away the remainder rounding down. In the end, our matrix representing $I$ which was $(N,N)$ in size has become a vector of of size $(N^2,1)$, and the element-wise product $|\mu|\mathcal{A}$ which was $(N,N)$ as well is now $(1,N^2)$. Simply multiplying these vectors in a matrix form computes the sum in (14), resulting in a $(1,1)$ output for $\tilde{I}(\phi)$.

So it is possible to use a matrix operation on a vectorized intensity distribution to recover a single azimuthal element of the Einstein Ring. In order to create a matrix model which can compute the entire azimuthal profile, we can use matrix concatenation. First a digression with a simple example, attempting to copy a vector. The $(2,2)$ identity matrix $\mathds{1}_2$ acts on a vector and preserves its entire structure. So, we can introduce the copy matrix
\begin{equation}
    \mathds{C}_2 = \begin{bmatrix} \mathds{1}_2 \\
    \mathds{1}_2
    \end{bmatrix},
\end{equation}
by vertically concatenating the identity matrix, to arrive at a matrix which acts on a vector by producing a copy concatenated with itself
\begin{equation}
    \mathds{C}_2  \begin{bmatrix}a & b
    \end{bmatrix} = \begin{bmatrix} 1 & 0 \\
    0 & 1 \\
    1 & 0 \\
    0 & 1
    \end{bmatrix} \begin{bmatrix}a & b
    \end{bmatrix} = \begin{bmatrix}a & b & a & b
    \end{bmatrix}.
\end{equation}
So if we would like to compute multiple azimuthal elements at angles $\phi_1, \phi_2, ..., \phi_m$ of the Einstein Ring in the matrix approximation, we can simply concatenate the matrices which perform each of the individual computations
\begin{equation}
    E = 
    \begin{bmatrix}
    |\mu|\mathcal{A}_{\phi_1} \\
    |\mu|\mathcal{A}_{\phi_2} \\
    \vdots \\
    |\mu|\mathcal{A}_{\phi_m} \\
    \end{bmatrix},
\end{equation}
which has a shape $(m,N^2)$ and it will act on a single vectorized intensity distribution to produce a vector of Einstein Ring measurements with shape $(m,1)$. 
\begin{figure}[h!]
    \centering
    \includegraphics[width=.8\textwidth]{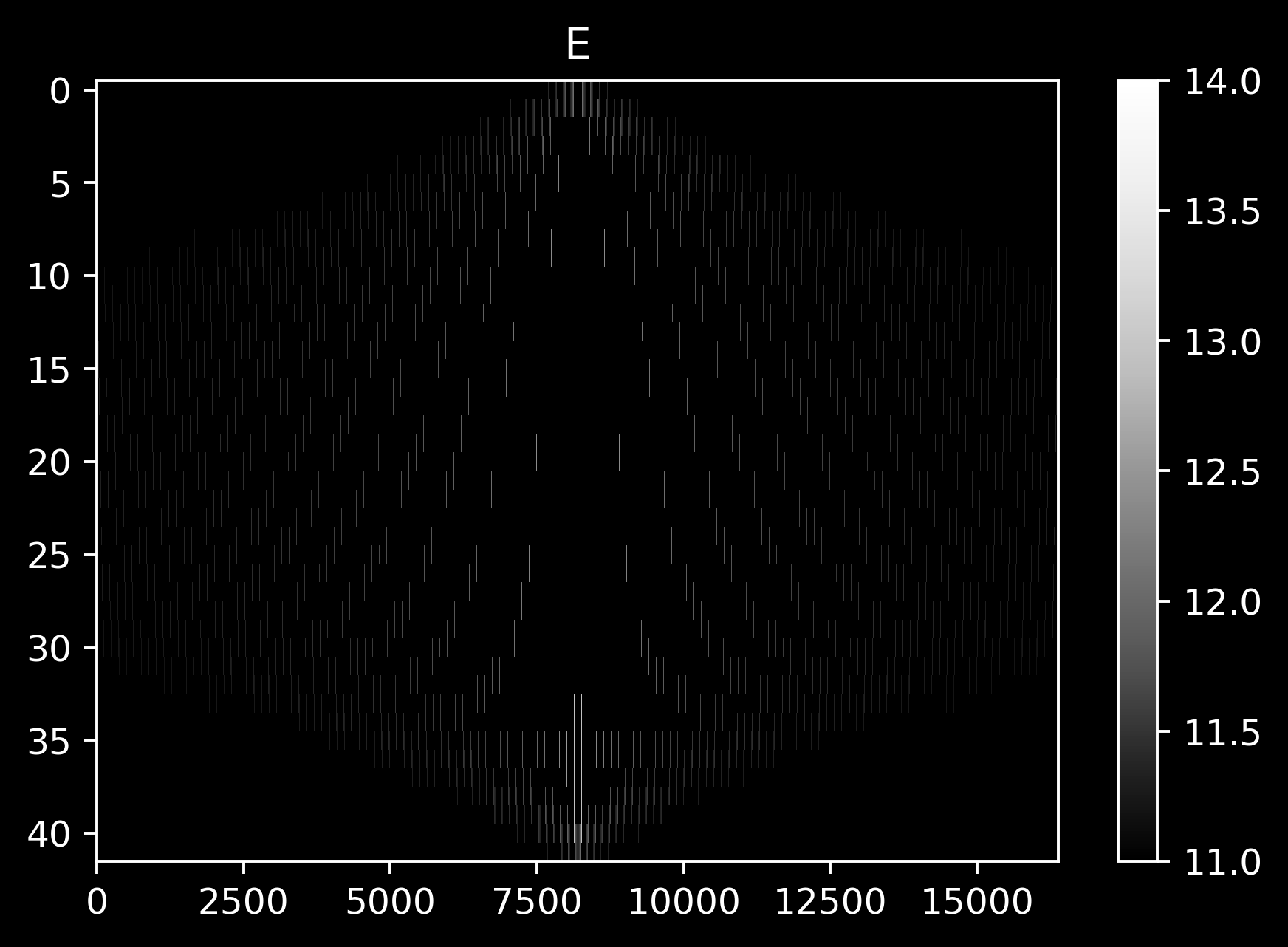}
    \caption{The matrix which computes the azimuthal profile of the Einstein Ring on a vectorized input. Color axis is logarithmic.}
    \label{fig:E}
\end{figure}

One might think it may be possible, with infinitely precise measurements of the Einstein Ring, reconstruction the input intensity distribution, but this is not the case. Because its spin-2 symmetry, the Einstein Ring profile does not contain all of the information present in the two dimensional intensity distribution. Because the Einstein Ring will appear identical if the input intensity distribution is rotated 180 degrees, this type of forward model cannot tell the difference between an input intensity distribution, and one that has been correspondingly symmetrized by taking the average between itself and itself rotated 180 degrees. This can be stated alternatively that an antisymmetric (about 180 degree rotations) intensity distribution is part of the null space of the matrix which forward models the Einstein Ring profile. If the two sectors on opposing sides of the planet have equal in magnitude but opposite in sign intensities, then they will sum to zero. In reality, negative intensities cannot exist, but any intensity distribution can be broken down into the sum of a symmetrized and antisymmetrized distribution, and a single infinite resolution measurement of the Einstein Ring cannot reconstruct the antisymmetric component.\footnote{It may be possible to reconstruct an intensity distribution which breaks this symmetry by its own virtue, such as one at optical wavelengths when the planet is illuminated on half of the disk in its apparent ``first quarter" or ``third quarter" lunar phase using this single measurement scheme, but such a reconstruction would be limited by the number of unique azimuthal elements which can be measured, which is not very large.}

In order to reconstruct the true intensity distribution, it is necessary to make multiple measurements of the Einstein Ring which have been slightly offset from one another. This can be done by means of sending the craft slightly off-axis from the planet and point deflector, which is geometrically equivalent to pointing our gravitational lens off center from the planet. This is accomplished in the matrix framework by an inventive scheme of zero-padding and shifting the input matrix with periodic boundary conditions. A simple example is demonstrated below for an image with $N=3$.

An arbitrary vectorized matrix may be shifted with periodic boundary conditions by multiplying by a single matrix $S$, which is a rearrangement of the identity matrix. The exact rearrangement depends on the unraveling scheme, shift axis, and number of pixels in the shift direction. But because the shift is linear, multiple shifts can be combined to produce arbitrary shifts by means of repeated multiplication of S. We begin by constructing the shift basis elements of a single pixel shift along each axis
\begin{eqnarray}
    \begin{bmatrix}
    a & b & c \\
    d & e & f \\
    g & h & i 
    \end{bmatrix}
    \xrightarrow{S_y}
    \begin{bmatrix}
    g & h & i \\
    a & b & c \\
    d & e & f
    \end{bmatrix}, \\
    \begin{bmatrix}
    a & b & c \\
    d & e & f \\
    g & h & i 
    \end{bmatrix}
    \xrightarrow{S_x}
    \begin{bmatrix}
    c & a & b \\
    f & d & e \\
    i & g & h
    \end{bmatrix}.
\end{eqnarray}
In the unraveling scheme described before, this may be accomplished by a $(9,9)$ matrix where the identity matrix has been regrouped within or using $(3,3)$ blocks in the following manner:
\begin{equation}
    S_y = \begin{bmatrix}
    &&&&&&1&& \\ 
    &&&&&&&1& \\ 
    &&&&&&&&1 \\ 
    1&&&&&&&& \\ 
    &1&&&&&&& \\ 
    &&1&&&&&& \\ 
    &&&1&&&&& \\ 
    &&&&1&&&& \\ 
    &&&&&1&&& \\ 
    \end{bmatrix} \\
\end{equation}
\begin{equation}
    S_x = \begin{bmatrix}
    &&1&&&&&& \\ 
    1&&&&&&&& \\ 
    &1&&&&&&& \\ 
    &&&&&1&&& \\ 
    &&&1&&&&& \\ 
    &&&&1&&&& \\ 
    &&&&&&&&1 \\ 
    &&&&&&1&& \\ 
    &&&&&&&1& \\ 
    \end{bmatrix}.
\end{equation}
This can be directly extended to shift matrices of size $(N,N)$ in analogous form
\begin{eqnarray}
    S_y = \begin{bmatrix}
    & & & \mathds{1}_N \\
    \mathds{1}_N & & & \\
    & \ddots & & \\
    & & \mathds{1}_N &
    \end{bmatrix} \\
    S_x = \begin{bmatrix}
    S_y\mathds{1}_N & \\
    & \ddots &\\
    & & S_y\mathds{1}_N
    \end{bmatrix}.
\end{eqnarray}
where each array is now $(N^2,N^2)$ and each sub-block of size $(N,N)$ is either all zeros, the identity matrix, or a shift thereof. By multiplying by these matrices each a certain number of times, it is possible to achieve arbitrary integer shifts. Defining the shift matrix
\begin{equation}
    S_{n_x,n_y} = S_{x}^{n_x}S_{y}^{n_y},
\end{equation}
the operation of a shift of $n_x,n_y$ along both axes can be applied to an arbitrary vectorized intensity distribution. If we wish to shift in the reverse direction (i.e. either $n_x < 0$ or $n_y < 0$, the periodic boundary conditions allow this by shifting $N+n_x$ or $N+n_y$ times, as both $S_x^N = S_y^N = \mathds{1}_N$.

\begin{figure}[h!]
    \centering
    \includegraphics[width=\textwidth]{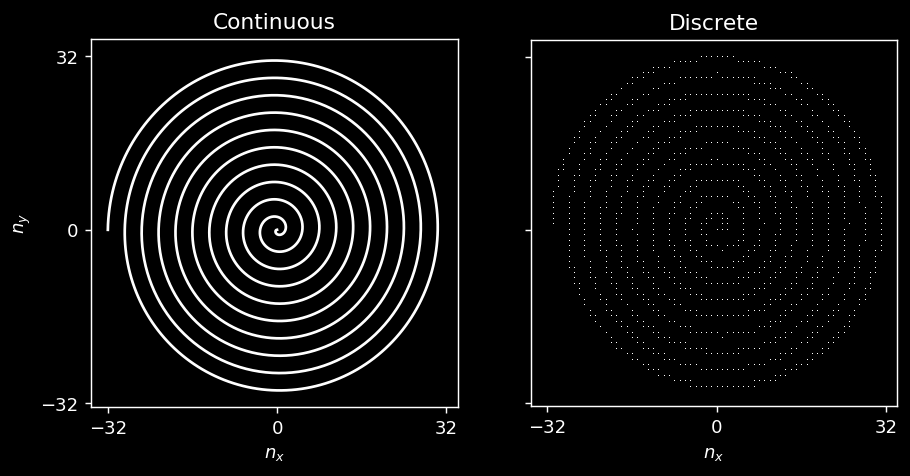}
    \caption{An example Archimedean spiral with 10 loops and its subsequent discretization into integer steps.}
    \label{fig:spiral}
\end{figure}

The exact question of which particular values of $n_x$ and $n_y$ are best remains open. The simplest approach would be to sample all values of $n_x$ and $n_y$, at least within the extent of the planet's disk, and this would correspond roughly to the strategies described previously\cite{Turyshev2020}, where the entire telescope acts as a single pixel detector in the reconstruction. However, this approach is potentially slower than necessary and could be exceedingly expensive in fuel costs. For our demonstration, we choose to parameterize a sparse spacecraft path in an Archimedean spiral \cite{archimedes} around the central axis of imaging. To make an Archimedean spiral with $N_\textrm{loops}$ number of loops, that extends to one quarter the size of the simulation box, so the maximum shift moves the edges of the planet right up against the edge of the box, we use the parametric equation
\begin{equation}
    \beta = \frac{\beta_\textrm{max}/2}{2\pi N_\textrm{loops}}\phi,
\end{equation}
where ($\beta,\phi$) are the polar coordinates, and $\phi \in (0,2\pi N_\textrm{loops})$. The $\xi$ and $\eta$ coordinates can be recovered using the standard relations $\xi = \beta \cos \phi$ and $\eta = \beta \sin \phi$, and compared to the values of the coordinates of the grid to produce a discrete realization of the spiral suitable for using with our integer shift matrix. The spiral and its discretization are plotted in Figure \ref{fig:spiral}.

The Archimedean spiral, or some other spiral, may prove to be an efficient path for sampling a circular subset of the two dimensional plane. A craft following such a trajectory may minimize its fuel expenditure by requiring the least instantaneous acceleration at each point, although a more detailed orbital mechanics calculation will be needed to verify this claim.

Now that all of those preliminaries have been established, it is finally possible to create the mega matrix that models all of the observations. By multiplying the shift matrix $S_{n_x,n_y}$ with the stack of matrices $E$ that computes the azimuthal profile (23), and stacking the various products for all of the shifts $(n_x,n_y)$ in our archimedan spiral together, we arrive at the end result:
\begin{equation}
    M_\textrm{Einstein} = \begin{bmatrix}
    E S_{n_{x_1},n_{y_1}} \\
    E S_{n_{x_2},n_{y_2}} \\
    \vdots
    \end{bmatrix},
\end{equation}
\begin{figure}[h!]
    \centering
    \includegraphics[width=\textwidth]{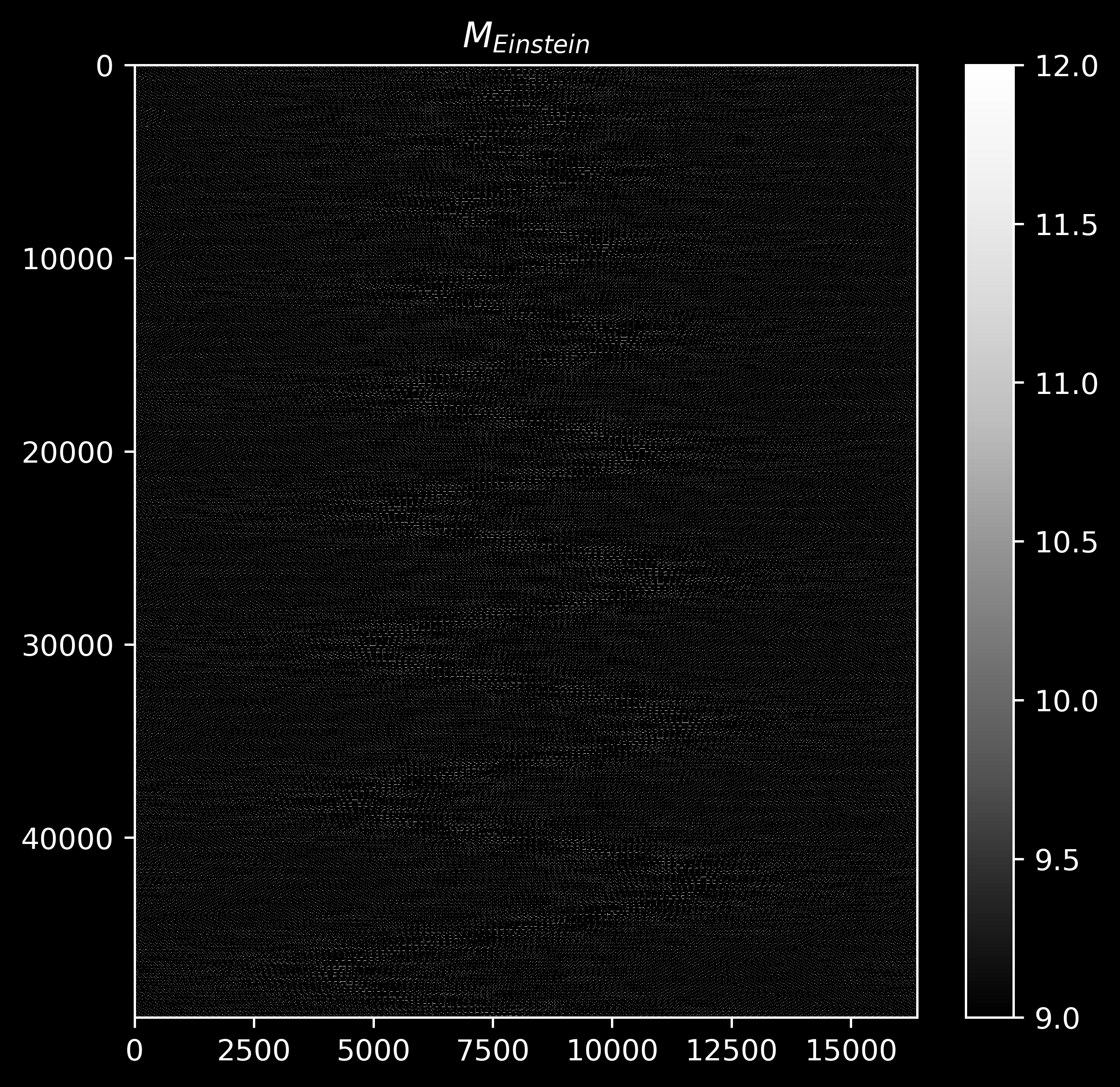}
    \caption{The Matrix which computes all of the Einstein Ring Observations. Color axis is logarithmic.}
    \label{fig:ME}
\end{figure}
which acts on an input vectorized intensity distribution and returns a vector of concatenated Einstein Ring profiles for the various observations. For comparison purposes we also construct the forward model for the total intensity only reconstructor which has been previously proposed. Such a forward model does not compute the Einstein Ring profile at each location but rather only its total intensity, which is equivalent to setting $\mathcal{A} = 1$ everywhere
\begin{equation}
    M_\textrm{total} = \begin{bmatrix}
    |\mu| S_{n_{x_1},n_{y_1}} \\
    |\mu| S_{n_{x_2},n_{y_2}} \\
    \vdots
    \end{bmatrix}.
\end{equation}
These matrices will in general be non-invertible, as there is no guarantee they will be square. However, the best unbiased estimator of the inverse for a linear problem is known as the pseudo-inverse and may be estimated by means of the singular value decomposition, which is a matrix factorization of the form
\begin{equation}
    M = U\Sigma V^*.
\end{equation}
Where $U$ and $V$ are square, real, orthonormal, and unitary matrices, and $\Sigma$ is a diagonal matrix containing the singular values. This is a useful method to decompose the linear transformation $M$, because $U$ and $V$ are unitary operators, they can be thought of as acting to rotate the basis elements of the space, while $\Sigma$ acts to stretch the rotated vector along the intermediary axis. This combination of rotate, stretch, derotate naturally allows one to find the pseudo-inverse, by the means of de-rotating, un-stretching, and re-rotating, written mathematically as
\begin{equation}
    M^+ = V\Sigma^{-1}U^*.
\end{equation}
The inverse of $\Sigma$ is simply the inverse of each of the diagonal elements, which is why numerically small singular values can cause the pseudo-inverse to be discontinuous, and many routines to find the pseudo-inverse will introduce a cutoff, below which the singular values are set to zero. This cutoff parameter is known as $rcond$, and is typically set to values near the machine epsilon, where numerical noise can become problematic. For our purposes, this default is a bit too small, but we find that values for $rcond$ in the range of $10^{-4}$ to $10^{-14}$ all behave well, with similar performance.

\begin{figure}[h!]
    \centering
    \includegraphics[width=\textwidth]{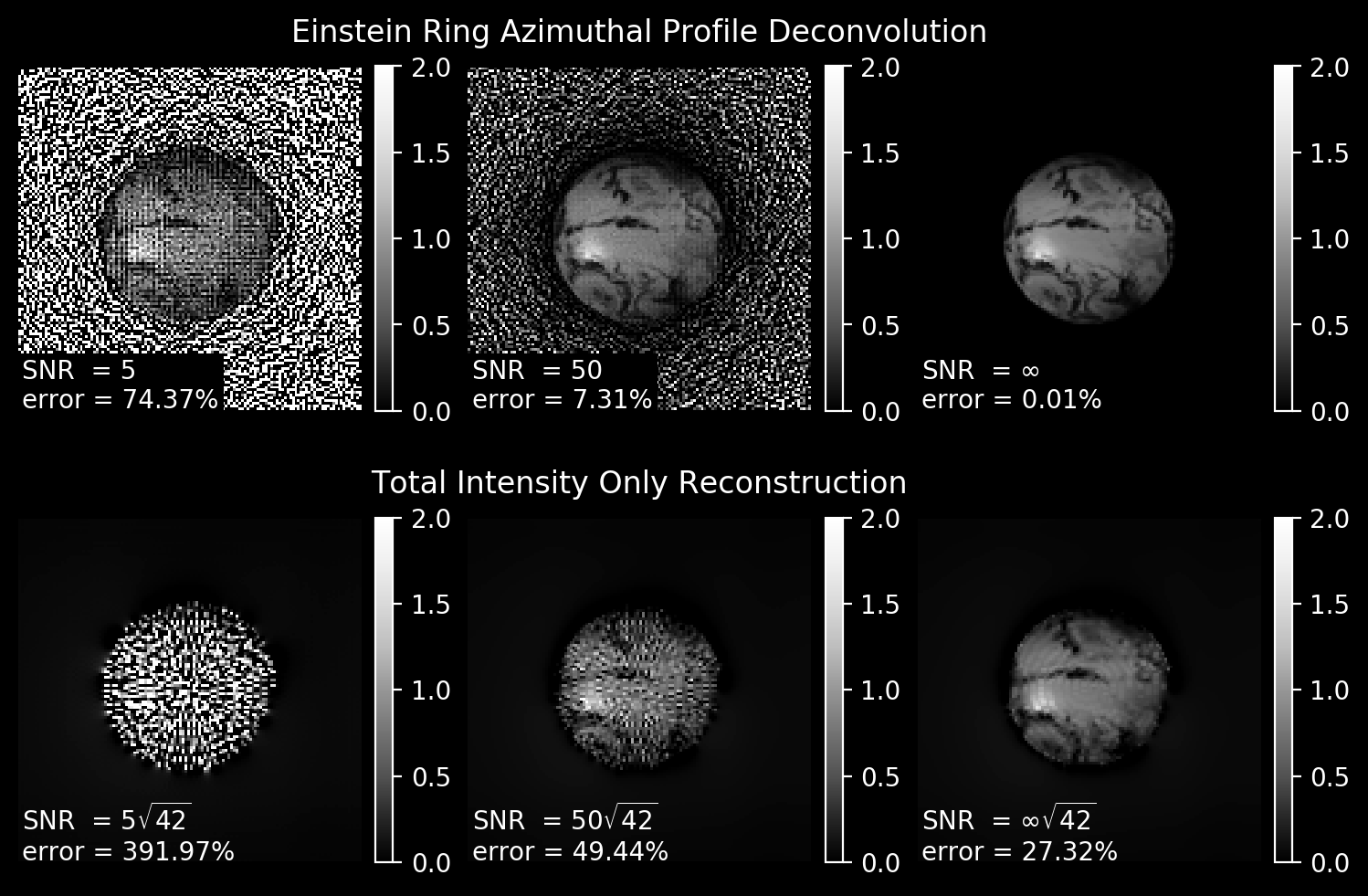}
    \caption{Comparison of the azimuthal deconvolution and total intensity reconstructors at various signal-to-noise ratios. The total intensity reconstructor SNR is boosted by a factor $\sqrt{42}$ so that each case corresponds to an equivalent integration time.}
    \label{fig:recon}
\end{figure}

To evaluate the relative performance of the azimuthal reconstructor to the total intensity reconstructor, we forward model an example intensity distribution, add measurement noise, and apply the pseudo-inverse. The measurement noise takes the form of a sample generated from a Gaussian distribution with zero mean and variance corresponding to the inverse of the SNR
\begin{equation}
    \hat{I} = M^+(MI + \mathcal{N}(0,1/\textrm{SNR})).
\end{equation}
To give the total intensity reconstructor a fair level of noise, we assume that the SNR on the total intensity is improved by a factor $\sqrt{42}$, which is the number of distinct azimuthal elements used. This puts the measurement between the two reconstructors on the same footing, as the total intensity reconstructor will be able to average down the noise by the square root of the number of samples it adds together, as if it had made the azimuthal measurements but decided not to use them. For both reconstruction paradigms, we evaluate the performance with an error metric
\begin{equation}
    \textrm{error} = \frac{\sqrt{\langle(I - \hat{I})^2 \rangle}}{\sqrt{\langle I^2 \rangle}},
\end{equation}
which is the fractional RMS reconstruction error. These are reported for the region corresponding to the planet in percent by multiplying by 100 along with the reconstructions for various SNRs in Figure \ref{fig:recon}.

By examining the relative performance it is apparent that using the additional azimuthal information at a set level of SNR for each resolution element in a sparse sampling of the image plane improves the reconstruction performance. Due to the sparse sampling, even at infinite SNR, the total intensity reconstructor has residuals corresponding to the spiral path the craft makes through the image plane. Since these points will be sampled at the corresponding time the craft passes through the point, the image will be mixed temporally, further complicating interpretations of the image. The azimuthal reconstructor could be useful to reduce the number of points needed to sample in the image plane, providing faster reconstructions that can better time resolve the changes on the planet's surface, which is useful to find the signature of continent's thermal response to its host star insolation.

\section*{Remarks}

\begin{figure}[h!]
    \centering
    \includegraphics[width=.5\textwidth]{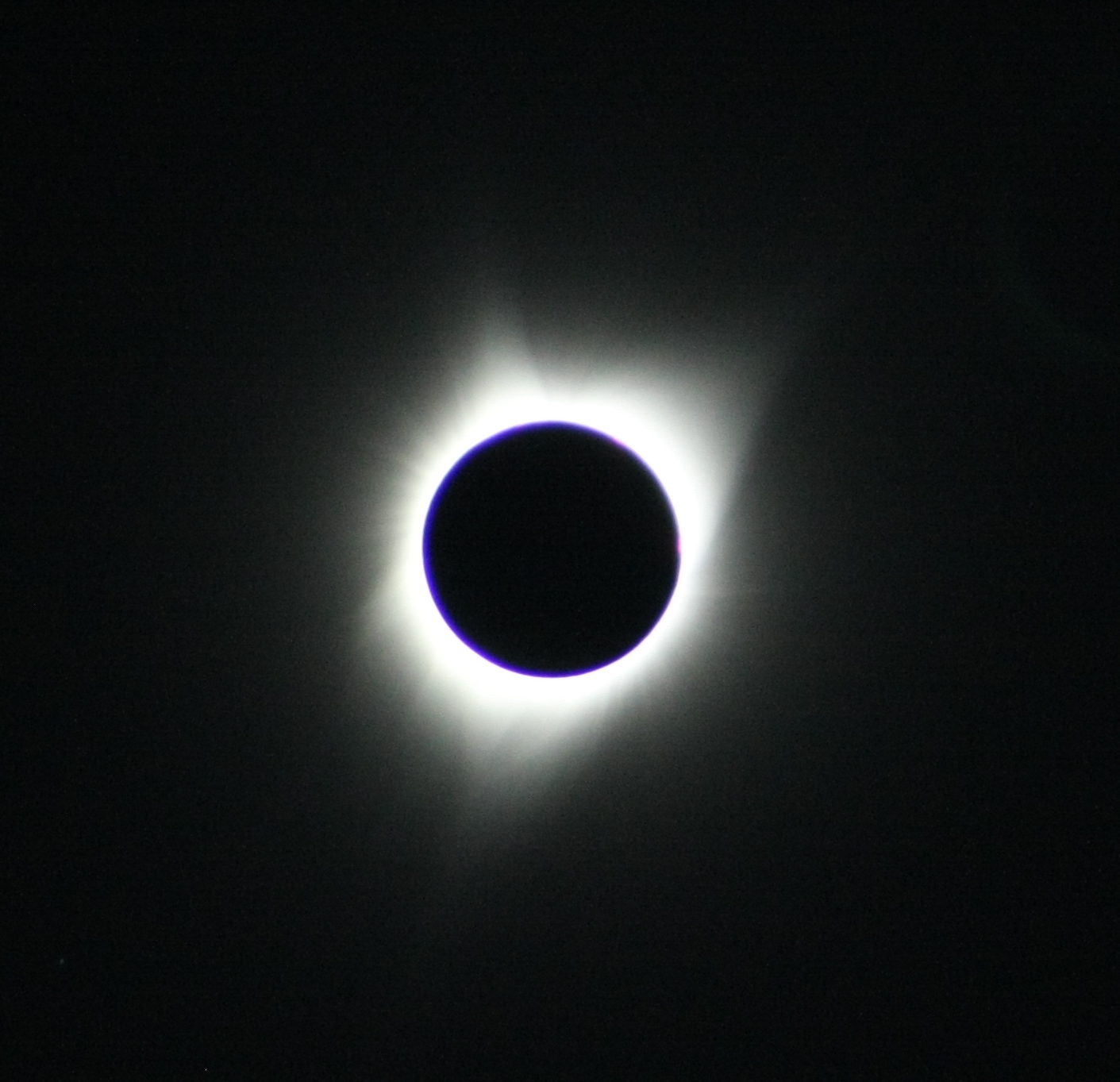}
    \caption{A view of the solar corona taken during the Total Solar Eclipse of August 21, 2017.}
    \label{fig:corona}
\end{figure}

To extract the signal from the Einstein Ring, extremely efficient suppression of solar noise contribution is required, such as an advanced coronagraph \cite{Lyot} or starshade \cite{Cash2006} which has been optimized for blocking sources with extended emission \cite{ferrari2009}, perhaps in addition to a secondary off-axis imager for reference subtraction \cite{Ruane2019} of the noise from the solar coronal emission, see the image in Figure \ref{fig:corona}.

A detailed instrument design and mission plan to resolve exocontinents would need to evaluate the requirements in detail to ensure that a reasonable level of SNR for each azimuthal resolution element could be acquired rapidly enough. Viewing a planetary system is by no means still life photography. Surface albedo changes on timescales of minutes and hours by rapidly evolving weather patterns, as well as on timescales of days and months as the planet rotates around its own axis as well as around its host star, modifying its apparent ``lunar phase" in wavelengths where illumination from the host star are dominant. Questions considering details such as optical transmission, throughput, quantum efficiency of detectors, would need to weighed against the magnification of the gravitational lens. Trade-offs between sensitivity and wavelength resolution in a spectrographic mode could be considered in the context of science questions, such as resolving photosynthetic vegetative regions with a red edge \cite{Seager2005}, or perhaps looking for extraterrestrial cities and their corresponding greenhouse gas emission \cite{Brindley2016} or from their artificial illumination during night \cite{Rosebrugh1935}.

The problem of reaching the focus of the solar gravitational lens at $\geq600$ AU has certainly not been solved yet either. At over four times further away from Earth than Voyager 1 has reached after 42 years of travel \cite{voyager}, it would be an incredible feat of space travel to reach that distance in a human lifetime. However, not all hope is lost. Ambitious projects like the Breakthrough Starshot \cite{PARKIN2018} program are working on powerful propulsion systems such as the laser-powered solar sail, which could overcome the inherent difficulties of the rocket equation to reach very high speeds. This doesn't completely eliminate the problem of fuel expenditure, as the craft will still have to accelerate to navigate the focal plane of the gravitational lens to sample different regions of the target. However, using the azimuthal variations in intensity could improve the efficiency of the source image reconstruction, saving fuel and time for the craft and mission. 

The last issue with this mission concept is its inherent one-off nature. At such an extreme heliocentric distance, even if it were possible to have the craft enter into a circular orbit, the repositioning time between targets is immense. The period corresponding to a single circular orbit is $T = 2\pi\sqrt{\frac{a^3}{GM}} \sim 15000$ years at $a = 600$ AU. Although it may be worth considering if a trajectory to acquire multiple planets colocated in a multi-planet system such as Trappist-1 \cite{gillon2017} is possible, it is immediately clear that a such a mission is very limited in its pointing capability.

All of these problems may be mitigated by simply acquiring a better gravitational lens. A solar mass black hole is an ideal candidate. At an orbital distance of $1$ AU, the black hole's Schwarzschild radius occupies a meager $4$ milliarcseconds of angular extent, while bringing the Einstein Ring into a much better view with an angular extent of $41$ arcseconds. This facilitates the measurement of additional azimuthal elements, which could further improve the reconstruction time. If the solar mass black hole is quiescent, with little accretion, it will be very dark, with only the Hawking Radiation \cite{hawking1975} emitted to contribute to the noise. The close-in orbit would improve the repositioning times by four orders of magnitude, allowing for acquisition of many targets across the whole sky. A single craft on a highly elliptical orbit could efficiently reconfigure it apoapsis with maneuvers at periapsis and greatly extend its scientific output. While it may be silly to think about acquiring a black hole like a lens you could buy from a store, there is justification to this speculation. In the dusk of our era of the universe, when the stars have all but burned themselves out, an advanced civilization may thrive by harnessing the energy of black holes. Using either the Penrose process  \cite{Penrose}, or the Blandford-Znajek process \cite{blandford}, an advanced generator of yet unknown form may shelter the last descendants of humanity when all else is nought. Such a civilization may use their host black hole as a lens, in analogous form, searching for any remaining source of light. And if they do, they may yet be first hand eyewitnesses to the ever decreasing circles of the fractally nested photon rings \cite{Johnson2020} it would form. And if that is not a reassuring end to the universe, I do not know what is.

\bibliographystyle{plain}
\bibliography{main}

\end{document}